# Doping dependence of the many-body effects along the nodal direction in the high-$T_c$ cuprate $(Bi,Pb)_2Sr_2CaCu_2O_8$

A. Koitzsch[1], S.V. Borisenko[1], A.A. Kordyuk[1,2], T.K. Kim[1], M. Knupfer[1], J. Fink[1], H. Berger[3], and R. Follath[4]

[1] *Leibniz-Institute for Solid State and Materials Research, IFW-Dresden, P.O. Box 270116, D-01171 Dresden, Germany*

[2] *Institute for Metal Physics of National Academy of Sciences of Ukraine, 03142 Kyiv, Ukraine*

[3] *Institute of Physics of Complex Matter, EPFL, CH-1015 Lausanne, Switzerland*

[4] *BESSY GmbH, Albert-Einstein-Strasse 15, 12489 Berlin, Germany*

**Abstract**: Angle-resolved photoemission spectroscopy (ARPES) is used to study the doping dependence of the lifetime and the mass renormalization of the low energy excitations in high-$T_c$ cuprate $(Bi,Pb)_2Sr_2CaCu_2O_8$ along the zone diagonal. We find a linear energy dependence of the scattering rate for the underdoped samples and a quadratic energy dependence for the overdoped case. The mass enhancement of the quasiparticles due to the many body effects at the Fermi energy is found to be in the order of 2 and the renormalization extends over a large energy range for both the normal and the superconducting state. The much discussed kink in the dispersion around 70 meV is interpreted as a small additional effect at low temperatures.

For the past 16 years it has remained an unsolved puzzle as to which microscopic process drives the evolution of Mott type insulating undoped cuprates to a metallic state with rather conventional properties at the highly overdoped side of the phase diagram. In particular, the phases in between the pseudogap state and the "strange metal" state with its anomalous transport properties [1] are not understood. Several normal state properties could be described within the marginal Fermi liquid (MFL) – phenomenology [2]. The more conventional behavior on the overdoped side seems to correspond to a Fermi liquid (FL) – but this has yet to be conclusively shown. Various concepts for the microscopic description of the phase diagram have been developed [3] but no consensus has been reached. All of these theories introduce certain types of many body interactions. These interactions affect the low energy excitations of the system. Namely, they cause a typical scattering rate and a mass renormalization which are both directly observable by angle-resolved photoemission spectroscopy (ARPES).

Firstly, we investigate the scattering rate as a function of energy, systematically, from underdoped to overdoped $(Bi,Pb)_2Sr_2CaCu_2O_8$ ((Bi,Pb)-2212)) and find compelling evidence for a crossover from MFL like behavior to FL-like behavior. We limit ourselves to the zone diagonal, which has been investigated previously by ARPES [4, 5, 6, 7]. However, these studies have concentrated almost exclusively on measurements on samples with optimal doping [7]. Information on the energy dependence of the scattering rate can also be obtained from optical [8] and Raman spectroscopy [9]. However, these methods inherently integrate over large portions of the Fermi surface (FS) and as they represent two-particle probes, comparison to ARPES as a one-particle probe is difficult.

Secondly, the many body interactions not only impose a finite lifetime of the low energy excitations but, necessarily, also a renormalization of their energy. In this context a "renormalization energy scale" at ca. 70 meV- usually referred to as "kink"- has been found previously by ARPES studies in the dispersion along the nodal direction of Bi-2212 [6, 10, 11, 12, 13]. The kink has been attributed to a coupling of the quasiparticles to phonons or to spin fluctuations. We show here that i) the dispersion at high temperatures can be fully understood with a pure MFL-like approach, ii) the actual kink develops at low temperatures and is an additional effect, iii) the magnitude of the kink-renormalization is weak throughout all doping levels compared to the overall magnitude of the many-body mass renormalization present in the system. This study on the many body effects along the nodal direction is complementary to our recent studies of the antinodal direction [14]. There we found evidence for strong coupling between the magnetic resonance mode observed in INS and electronic states below $T_c$ around the M-point which also leads to the appearance of kinks in the dispersion. We discuss the nature of the nodal kinks in light of the characteristics of the antinodal kinks.

The ARPES experiments were carried out using radiation from the U125/1-PGM beam line and an angle multiplexing photoemission spectrometer (SCIENTA SES 100) at the BESSY synchrotron radiation facility. The spectra were recorded using excitation energies h$\nu$=25-27 eV with a total energy resolution of 10 meV. The momentum resolution was set to 0.01 Å$^{-1}$ parallel to $(0,0)-(\pi,\pi)$ and 0.02 Å$^{-1}$ perpendicular to this direction. Measurements have been performed on $(5\times1)$ superstructure free high quality single crystals of (Bi, Pb)-2212 with varying doping levels ranging from underdoped with T$_c$=76K (UD76) through optimally doped (OP89) with T$_c$=89K to overdoped (T$_c$=61K, OD61). The dopant concentration was calculated from the empirical equation of T$_c$ versus the dopant concentration [15]. Previously, for these compounds, the concentration values have been derived from the volume of the FS [16]. It has been found, that both methods were in agreement.

Figure 1 introduces the subject of our measurements. In Fig. 1a the FS of (Bi, Pb)-2212 is shown schematically. It consists of the typical bilayer split barrels located around the $(\pi,\pi)$ points of the Brillouin zone. At the crossing point of the diagonal and the FS, the bilayer splitting vanishes and also the superconducting order parameter has a node here. We performed several cuts (thin black lines) in the vicinity of the suspected nodal direction to identify this direction precisely. We then applied Lorentzian fits to the constant energy slices (Momentum Distribution Curves, MDC) of the two-dimensional dataset (Fig. 1b). The MDC at the Fermi energy is presented in Fig. 1d along with the fit. The position of the maximum gives us the dispersion and the width gives the lifetime of the excitation. For reference a cut at constant momentum (k$_F$) resulting in a Energy Distribution Curve (EDC) is presented in Fig. 1c.

The proper quantity used to describe many body interactions is the complex self energy $\Sigma$. The real part Re $\Sigma$ describes the energy renormalization of an excitation, the imaginary part describes the scattering rate. Im $\Sigma$ can be directly expressed in terms of the Full Width Half Maximum (FWHM) of the measured peaks: $\mathrm{Im}\Sigma = v_0/2 \times FWHM$, where v$_0$ is the bare Fermi velocity of the noninteracting system. A physically meaningful determination of v$_0$ is crucial for a quantitative analysis of Im $\Sigma$. We use here a value of v$_0$=4 eVÅ as derived from the detailed analysis of the anisotropic plasmon dispersion in Bi-2212 [17], which is determined by the mean unrenormalized Fermi velocity. The result is also consistent with LDA bandstructure calculations [18]. Both methods are sensitive to the bare particle behavior of the electronic system. Another confirmation of $v_0$=4 eVÅ comes from the analysis of photoemission data itself: imaginary and real part of $\Sigma$ do not represent independent quantities. For instance enhanced scattering rates enforce ultimately decreasing bandwidth, i.e. enhancement of Re $\Sigma$. It can be shown that Re $\Sigma$ and Im $\Sigma$ fulfill Kramers Kronig relation. Making use of this relation Im $\Sigma$ can be inverted to Re $\Sigma$ and the bare dispersion is obtained by subtracting Re $\Sigma$ from the measured dispersion [19]. Our value for $v_0$ and the procedure to extract it are different from previous studies [10, 11, 12]. There, $v_0$ has been extracted by the ad hoc assumption that the bare dispersion matches the measured dispersion at the highest binding energies measured. However, we believe that our method is physically more justified.

In Fig. 2 Im $\Sigma$ is presented as a function of energy for various doping levels. The most underdoped sample shows a nearly linear dependence of Im $\Sigma$ on energy and correspondingly. With increasing hole doping the dependence becomes more rounded and a drop of Im $\Sigma$ below ca. 80 meV develops and is most pronounced for the optimal doped case. The drop in Im $\Sigma$ vanishes for higher temperatures (not shown) in agreement with previous results [5, 6]. Further doping seems to shift the energy upward where the drop occurs (OD81). At higher doping levels a parabolic shape over the whole energy range is observed (OD77, OD61) [20]. A previous study of overdoped Bi-2212 reported a linear dependence [7]. We believe that our data are more conclusive in terms of the signal to noise level of the spectra and the greater number of different samples presented. All of the curves presented in Fig. 2 have a zero-energy offset, which we attribute to the finite momentum resolution and elastic defect scattering.

The continuous evolution from a linear to a parabolic line shape of the *energy* dependence in going from under to overdoping suggests a transition from a MFL-like regime to a FL regime. A similar behavior was found in transport [1] and ARPES [4, 7] experiments for the scattering rate at the Fermi level as a function of *temperature*. These two aspects show conclusively that the fundamental many body interactions in cuprates change with doping. Im $\Sigma$ and hence the many body effects are comparable to the binding energies of the excitations. We therefore expect a large renormalization in the dispersion of the low energy excitations and this is now discussed.

The dispersion along the nodal direction for several doping levels is given in Fig. 3a. The dispersions are similar for all the doping levels presented. Small differences exist around the kink energy but appear to be subtle and will not be discussed here. On one hand it seems sur-prising that only small systematic changes with doping are observed given the systematic changes in Im Σ. On the other hand the absolute differences among the doping levels for Im Σ are also small and lead only to small changes in Re Σ which are unlikely to be detected in the dispersions. The dashed straight line represents an approximation to the bare dispersion with a slope of $v_0$=4eVÅ as discussed above. We emphasize that the bare dispersion is not only a theoretical limit but a well defined and observable entity [17]. In Fig.3 b we show a comparison between T=30 K and T=300 K for the optimally doped sample. The overall temperature dependence is weak and agrees with previous investigations [6, 11, 12]. The real part of Σ is the difference between the measured dispersion and the bare dispersion. This is shown for the underdoped sample for T=30 K and T=200 K in Fig3 c and d. The dashed lines represent a fit to the MFL relationship [2]:

$$\mathrm{Re}\,\Sigma = \alpha\omega\ln\left(\frac{\sqrt{\omega^2+\pi^2T^2}}{\omega_C}\right),$$

where the constant $\alpha$ and the cutoff frequency $\omega_C$ were allowed to vary freely for the T=200 K spectrum. The fit is in excellent agreement with the data for $\alpha$=0.22 and $\omega_C$=1.59 eV. Using the same parameters for the low temperature spectrum in Fig 3 c results in a disagreement for values below ca. 60 meV (i.e. at the kink energy scale). Allowing the parameters to vary freely does not result in a more accurate fit (not shown).

We conclude from this observation that for underdoping the normal state is well described by the MFL-like approach. The dispersion has a curvature which mimics a smooth kink. At low temperatures the actual kink manifests itself as a deviation from the MFL-like fit, and could be related to the onset of superconductivity. The magnitude of the renormalization described by the MFL exceeds the magnitude of the kink renormalization significantly. Using $\lambda = \frac{d(\mathrm{Re}\Sigma)}{dE}(E=0)$ we obtain $\lambda = 0.8 \pm 0.1$ for the coupling constant $\lambda$ in the normal state which is similar to values recently reported for the antinodal direction [14]. This indicates that the many body effects in the normal state of (Bi,Pb)-2212 are approximately isotropic and extend over a wide energy range (in the order of $\omega_c$). Whereas it is well known that the electron-electron interaction is responsible for the characteristics of a FL a variety of proposals exists for the microscopic origin of the MFL phenomenology. From one point of view there exists a hidden order parameter influencing the low energy excitations [21]. Another proposal stresses the importance of the antiferromagnetic phase resulting in a coupling of the charge carriers to overdamped spin fluctuations [22, 23]. In the light of our recent results [24] we find the latter explanation appealing but further investigations are clearly needed.

We now discuss the origin of the kink at low temperatures. From our point of view, the absence of the kink at T=200 K makes an interpretation in terms of a coupling to phonons questionable. Rather the results suggest a coupling to the resonance mode observed in inelastic neutron scattering at temperatures below $T_c$ [22, 25, 26]. While the impact on the electronic states of this mode is maximal at the M points [13, 14] remnants could still be observable along the nodal direction [22]. Since the effect is small and could even be masked by thermal broadening it is more difficult to collect a similar set of evidence as we have done in [14]. Nevertheless, this scheme offers a natural and unified explanation for the observed mass enhancement throughout the Brillouin zone, which is difficult to obtain assuming a phonon coupling. Such a coupling should be more isotropic and independent of doping. It is, however, possible that both effects are present with different weight depending on the k-space location.

In summary we have shown that the scattering rate crosses in continuous manner from a more linear dependence on energy to a more quadratic dependence with overdoping, which represents a crossover from a MFL-like regime to a FL. We find the real part of the self energy extracted from the dispersion of an underdoped sample in the normal state in agreement with a MFL-like approach. The mass renormalization extends to high energies. At low temperatures a kink is observed around 70 meV which could be due to the coupling of the quasiparticles to remnants of the magnetic resonance mode.

We acknowledge stimulating discussions with C.M. Varma and K. Matho. This work was financially supported by the DFG (Fi 439/10-1). HB is grateful to the Fonds National Suisse de la Recherche Scientifique. The sample preparation in Lausanne was supported by the NCCR research pool MaNEP of the Swiss NSF.

.

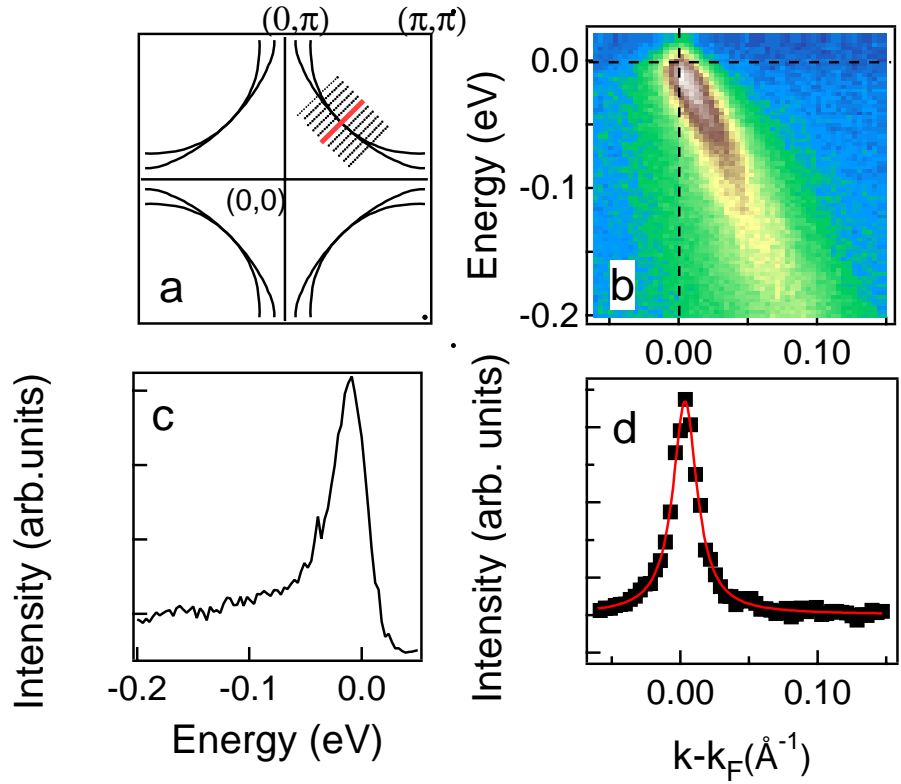

Figure 1

a) Fermi surface of (Bi,Pb)-2212 (schematically) with the typical bilayer split barrels around the $(\pi,\pi)$ points. The middle line indicates the position along the nodal direction where the Energy Distribution Map (EDM) in (b) was taken. The row of stacked thin lines indicate the positions of EDM's taken to identify the nodal direction. b) EDM taken at 30 K for an underdoped sample along $(0,0)$-$(\pi,\pi)$. c) EDC along the vertical dashed line in (b). d) MDC at the Fermi energy. The solid line is simple Lorentzian fit.

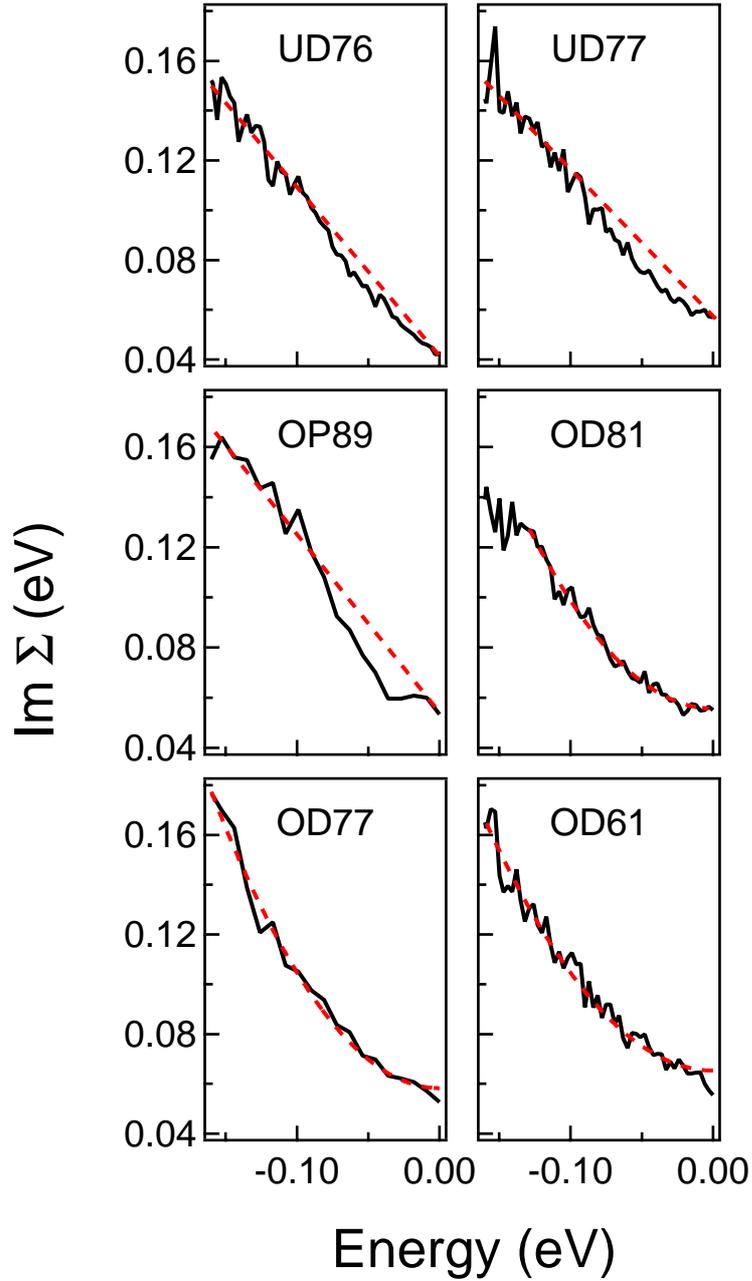

Figure 2

Imaginary part of the self energy Σ as a function of binding energy extracted from the Energy Distribution Map's (EDM) at T=30K for various doping levels. Note the change from a more linear behavior for underdoping to a parabola for overdoping. The dashed straight lines for the underdoped samples (UD76, UD77) and the optimally doped (OP89) are guides to the eye, showing the approximate linearity for underdoping and how the drop below ca. 80 meV develops for optimal doping. The dashed lines for the overdoped samples are fits to a parabola.

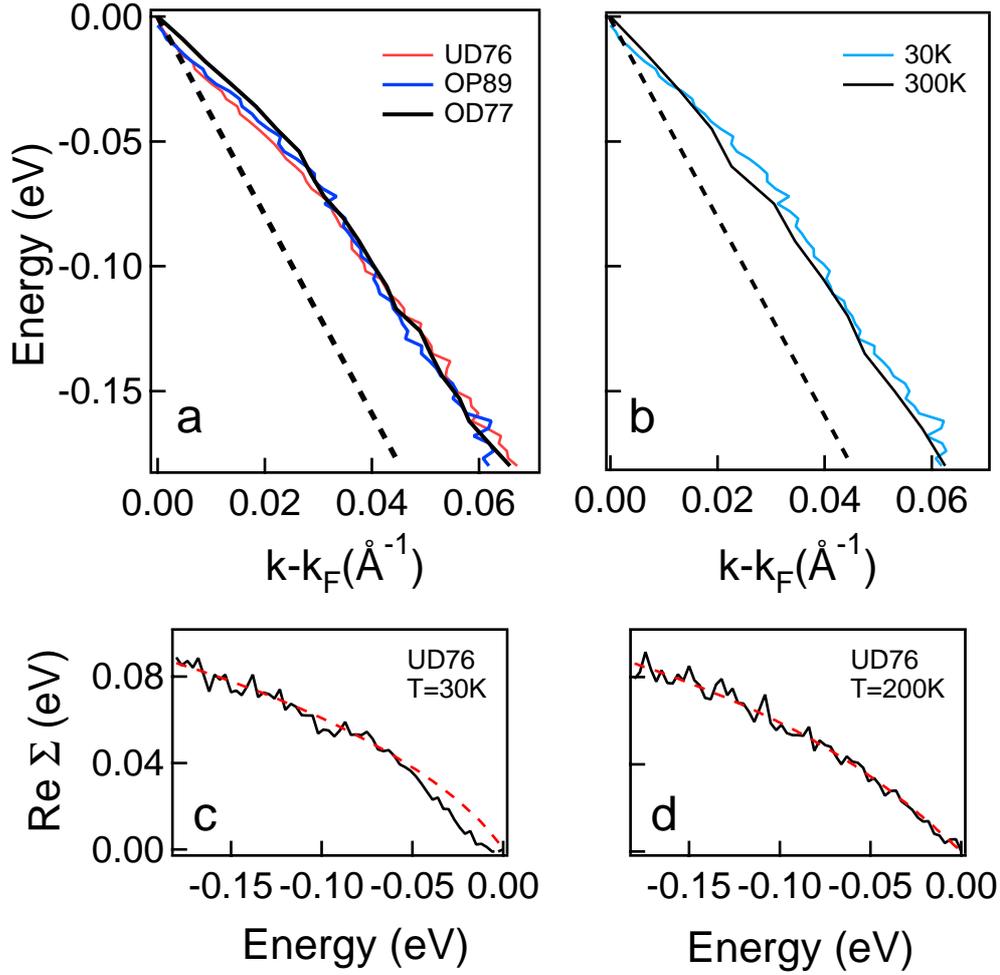

Figure 3

a) Dispersions for under, optimally and overdoped samples at T=30 K. The curves have been shifted to match the same $k_F$. The dashed line represents the bare dispersion as explained in the text. b) Comparison of the T=30 K and T=300 K dispersions for the optimally doped sample. c), d) Real part of $\Sigma$ evaluated as the difference between the dispersion and the straight line in (a) for the underdoped sample. The dashed line is a fit to a Marginal Fermi Liquid form (see text). Note the good agreement for T=200 K and the deviations at the kink energy scale for T=30 K.